\documentstyle[preprint,revtex]{aps}
\def\ln{{\rm ln }}
\def\li{{\rm Li}_2}
\begin{document}
\draft
\centerline{\hfill NUHEP--TH--92--11}
\pagestyle{empty}
\begin{title}
Anomalous Radiative Decay of Heavy Higgs Boson\\
\end{title}
\author{Tzu Chiang Yuan\\
\quad\\ \quad\\}
\begin{instit}
Department of Physics and Astronomy,
Northwestern University, Evanston, IL 60208 \\
\end{instit}
\vfill
\vfill
\begin{abstract}
The radiative decay width of a heavy Higgs boson
$H \rightarrow W^+W^-\gamma$ for a {\it hard} photon is calculated
in the Standard Model and its extension with anomalous $\gamma WW$
couplings. Its dependence on the Higgs mass, the two unknown anomalous
couplings, and the photon energy cutoff are studied in detail. We show that
this radiative decay of a heavy Higgs is not very sensitive to
a wide range of the anomalous couplings compared to the Standard Model
result.
\end{abstract}
\vfill
%
%
\pagestyle{plain}

\section{INTRODUCTION}

If the Standard Model (SM) scalar Higgs boson is heavier than twice the W-
or Z-boson masses, it will decay predominantly into the two gauge bosons.
Hunting for such a heavy Higgs is one of the primary goal at the
future hadron colliders like SSC or LHC\cite{higgshunter}.
The decay rate for $H \rightarrow W^+W^-$ is given by
\begin{equation}
\Gamma (H \rightarrow W^+W^-) \; = \; {\alpha h \over 16\sin^2\theta_w}
                               {(1-r^2)^{1 \over 2} \over r^2}
                               (3r^4 - 4r^2 +4) \quad ,
\end{equation}
where $r = 2w/h$
($w$ and $h$ denote the W-boson and Higgs masses respectively).
This partial width increases
monotonically with the Higgs mass and eventually violates the unitarity
bound -- this indicates the heavy Higgs boson couples strongly
with the longitudinal component of the gauge boson. Similar feature
holds for $H \rightarrow ZZ$. The radiative
decay $H \rightarrow W^+W^-\gamma$ for a ${\it hard}$ photon with energy
$E_{\gamma} \ge  5$ GeV has also
been considered in the SM\cite{dicusetal}. Despite the branching ratio
\begin{equation}
R_{hard} = \Gamma (H \rightarrow W^+W^-\gamma)/\Gamma (H \rightarrow W^+W^-)
\end{equation}
is only about several percent, it grows with the Higgs mass.
(We note that the multi-soft photons contribution of this process
as well as the full SM one-loop electroweak corrections to
$H \rightarrow W^+W^-$ have been thoroughly studied recently in
\cite{kniehl}.)
In this paper we extend this previous work of \cite{dicusetal}
by including the anomalous $\gamma WW$ couplings which are allowed
by the discrete T, C, and P invariance and consistent with
electromagnetism. These new contributions to the $\gamma WW$
vertex can be induced at the one-loop level in SM
or its various extension (for example,
the two-Higgs-doublets model\cite{rizzo}).
Thus the radiative decay mode can be used to probe either
the SM at the quantum level or new physics
(for example compositeness, supersymmetry, extended Higgs sector, ...)
or both! In the next section, we present the matrix
element of the process $H \rightarrow W^+W^-\gamma$.
We then study the Higgs mass dependence of the branching ratio $R_{hard}$
for a {\it hard} photon in a wide class of model parametrized by the anomalous
couplings in section 3. Analytic formulas for
the decay rate are relegated to an Appendix.

\section{DECAY RATE OF $\Gamma (H \rightarrow W^+W^-\gamma)$}

The general $\gamma WW$ couplings that are allowed by electromagnetism
and the discrete T, C, and P invariance have been
written down in \cite{gaemers},
\begin{equation}
{\cal L}_{\gamma WW} \; = \; -ie \left[ (W_{\mu\nu}^\dagger W^\mu A^\nu
                                    - W^\dagger_\mu A_\nu W^{\mu\nu} )
              + \kappa W^\dagger_\mu W_\nu F^{\mu\nu}
              + {\lambda \over w^2}
               W^\dagger_{\lambda\mu}W^{\mu}\,_{\nu}F^{\nu\lambda} \right]
\; \; .
\end{equation}
The anomalous couplings $\kappa$ and $\lambda$ are related to the
magnetic moment $\mu_W$ and the electric quadrupole moment
$Q_W$ of the W-boson defined by
\begin{equation}
\mu_W \; = \; {e \over 2w}(1+\kappa+\lambda) \; \; , \; \; \; \; \;
Q_W   \; = \; - {e \over w^2}(\kappa - \lambda) \; \; .
\end{equation}
In SM, $\delta \kappa \equiv \kappa -1 = 0$ and $\lambda = 0$ at tree level.
There are two Feynman diagrams contribute to the process
$H \rightarrow W^+(k_1)W^-(k_2)\gamma(q)$, since the photon can couple either
to $W^+$ or $W^-$. The decay rate
can be calculated readily and is given by (we follow some of the notations
of Ref.\cite{dicusetal})
\begin{equation}
\Gamma(H \rightarrow W^+W^-\gamma) \; = \;
              {\alpha^2 h \over 16\pi\sin^2\theta_w}
\int^{1-r^2}_{y} dx \int^{x_+^{max}}_{x_+^{min}} dx_+ W \quad ,
\end{equation}
where $x  = {2E_{\gamma} \over h}$ and
$x_+  =  {2E_{W^+} \over h}$ are the rescaled energies of the photon and
the W$^+$-boson respectively.  The integration range of $x_+$ is
\begin{equation}
x_+^{max,min} \; = \; 1 - {x \over 2} \pm {x \over 2(1-x)}R(x,r) \; \; \; ,
\; \; R(x,r) \; = \; \sqrt{(1-x)(1-x-r^2)} \; .
\end{equation}
Due to the infrared divergencies associated with emission of soft photons,
we cutoff the lower end of the $x$-integral at
$y = {2E_{\gamma}^{min} \over h}$. In terms of these rescaled variables,
we have
\begin{equation}
q \cdot k_1 \; = \; - {h^2 \over 2}(1-x-x_+) \; \; , \; \;
q \cdot k_2 \; = \;   {h^2 \over 2}(1-x_+) \; \;  , \; \;
k_1 \cdot k_2 \; = \;  {h^2 \over 2}(1-x-{r^2 \over 2}) \; \; .
\end{equation}
The matrix element squared is given by
\begin{equation}
W \; = \; W_{SM} + \lambda W_{\lambda} + \delta \kappa W_{\delta\kappa}
          + \lambda^2 W_{\lambda^2} + (\delta \kappa)^2 W_{\delta\kappa^2}
          + \lambda \delta \kappa W_{\lambda\delta\kappa} \;\; ,
\end{equation}
where
\begin{equation}
\begin{array}{cl}
W_{SM}  \; =  \; &  + 2 + 4{k_1 \cdot k_2 \over w^2}
           + \left( 2 + {(k_1 \cdot k_2)^2 \over w^4} \right)
           \left[ 2{w^2 k_1 \cdot k_2 \over q \cdot k_1 q \cdot k_2}
                 - {w^4 \over (q \cdot k_1)^2}
                 - {w^4 \over (q \cdot k_2)^2} \right]
 \\
  &  + {2 \over q \cdot k_1} \left[ q \cdot k_2 - k_1 \cdot k_2
        + {1 \over w^2}q \cdot k_2 k_1 \cdot k_2 + {2 \over w^2}
             (k_1 \cdot k_2)^2 \right]
 \\
  &  + {2 \over q \cdot k_2} \left[ q \cdot k_1 - k_1 \cdot k_2
        + {1 \over w^2}q \cdot k_1 k_1 \cdot k_2 + {2 \over w^2}
             (k_1 \cdot k_2)^2 \right]
 \\
  &  + {1 \over (q \cdot k_1)^2} \left[ (q \cdot k_2)^2 -
                 2 q \cdot k_2 k_1 \cdot k_2 \right]
 \\
  &  + {1 \over (q \cdot k_2)^2} \left[ (q \cdot k_1)^2 -
                 2 q \cdot k_1 k_1 \cdot k_2 \right] \; \; ,
 \\
\end{array}
\end{equation}
\begin{equation}
\begin{array}{cl}
W_{\lambda}  \; =  & +4 + {2 \over w^2}
            \left[ q \cdot k_1 + q \cdot k_2 \right]
 	+ 2{q \cdot k_2 \over q \cdot k_1}
        \left[ 2 + {1 \over w^2} (q \cdot k_2 + k_1 \cdot k_2) \right] \\
    &	+ 2{q \cdot k_1 \over q \cdot k_2}
            \left[ 2 + {1 \over w^2} (q \cdot k_1 + k_1 \cdot k_2) \right]
			\; \; ,
 \\
\end{array}
\end{equation}
\begin{equation}
\begin{array}{cl}
W_{\delta\kappa}  \; =  &  + 8 - {2 \over w^2}
            \left[ q \cdot k_1 + q \cdot k_2 \right]
     	+ 2 \left[ {(q \cdot k_1)^2 \over (q \cdot k_2)^2}
	+ {(q \cdot k_2)^2 \over (q \cdot k_1)^2} \right]
 \\
&   	+ 2{q \cdot k_2 \over q \cdot k_1}
            \left[ 2 - {1 \over w^2} (q \cdot k_2 + k_1 \cdot k_2) \right]
    	+ 2{q \cdot k_1 \over q \cdot k_2}
            \left[ 2 - {1 \over w^2} (q \cdot k_1 + k_1 \cdot k_2) \right]
		\; \; ,
 \\
\end{array}
\end{equation}
\begin{equation}
\begin{array}{cl}
W_{\lambda^2,\delta\kappa^2}  \;
          =  &  + {3 \over 2} + {(k_1 \cdot k_2)^2 \over w^4}
      \pm {2 \over w^2} \left( 1 - {k_1 \cdot k_2 \over 2w^2} \right)
      \left[ q \cdot k_1 + q \cdot k_2 \right]
 \\
&  + \left( 1 - {k_1 \cdot k_2 \over 2w^2} \right)
        \left[ {q \cdot k_2 \over q \cdot k_1}	 +
		{q \cdot k_1 \over q \cdot k_2}	 \right]
   + {1 \over 4} \left[ {(q \cdot k_2)^2 \over (q \cdot k_1)^2}	 +
		{(q \cdot k_1)^2 \over (q \cdot k_2)^2}	 \right]
 \\
&   \pm {1 \over w^2} \left[ {(q \cdot k_2)^2 \over q \cdot k_1} +
		{(q \cdot k_1)^2 \over q \cdot k_2}	 \right]
   + {1 \over 2w^4} \left[ (q \cdot k_1)^2 + (q \cdot k_2)^2	 \right]
	\; \; , \\
\end{array}
\end{equation}
and
\begin{equation}
W_{\lambda\delta\kappa}  \; =  \; W_{\lambda^2} + W_{\delta\kappa^2}
   - {2 \over w^4} \left[ (q \cdot k_1)^2 + (q \cdot k_2)^2	 \right]
\; \; .
\end{equation}
The SM result of $W_{SM}$ agrees with Ref.\cite{dicusetal}.  Our calculation
was performed in the unitary gauge. Noteworthy, for the process
that we are interested in, the $\lambda$-term of the
anomalous couplings in Eq.(3) do not contribute to the longitudinal piece
of the W-boson propagator. The above results give us the branching ratio
\begin{equation}
R_{hard} \;  = \;
   {\alpha \over \pi }r^2(1-r^2)^{-{1 \over 2}}(3r^4-4r^2+4)^{-1}
   \int^{1-r^2}_{y} dx  \int^{x_+^{max}}_{x_+^{min}} dx_+ W \quad .
\end{equation}
All the double integrals in Eq.(14) can be done analytically. The final
formulas are tedious and not illuminative, we therefore
relegate them to the Appendix.

\section{DISCUSSIONS}

Previously, a very weak experimental limit on $\kappa$
($-73.5 \le \kappa \le 37$ with 90 \% CL) has been derived\cite{grotchrobinett}
from PEP and PETRA
by studying the process $e^+ e^- \rightarrow \gamma \nu \bar \nu$.
Recently, more stringent limits of
$-3.5 < \kappa < 5.9$ (for $\lambda=0$) and
$-3.6 < \lambda < 3.5$ (for $\kappa=0$) with 95 \% CL were obtained
from the study of the process
$\bar p p \rightarrow e \nu \gamma + X$
by the UA2 Collaboration\cite{UA2}.
These limits are of course agree well with the SM tree level prediction.
Nevertheless, they are expected to be improved considerably at the TEVATRON
in the near future\cite{TEVATRON}. More accurate
measurements on the anomalous couplings $\vert \delta \kappa \vert$ and
$\vert \lambda \vert$ at the level of $\sim$ 0.1 -- 0.2
are expected at LEP II by studying the process
$e^+e^- \rightarrow W^+W^-$\cite{gaemers}. Also,
several recent studies\cite{HERA}
of the process $e^{\pm} p \rightarrow \nu \gamma + X$  conclude a
somewhat less sensitivity of the anomalous couplings at HERA.

Without referring to any particular values for the anomalous couplings in
any specific models, we are free to vary their magnitudes that are consistent
with the present UA2 experimental constraints.
In Figures (1a) and (1b), we plot the ratio $R_{hard}$ as function
of the Higgs mass with a photon energy cutoff $E_{\gamma}^{min} = 10$ GeV for
$(\delta \kappa , \lambda) = (\pm 0.5 , \pm 0.5)$ and $(\pm 1, \pm 1)$
respectively. The SM contribution is also presented for comparison. One
can see that in SM the branching ratio is less than 6 percent for the entire
range of the Higgs mass that we are interested in
(from 200 GeV to 1 TeV). The anomalous contributions
are not significant unless the magnitude of the anomalous
couplings $\delta \kappa$ and $\lambda$ are significantly larger
than 1. $R_{hard}$ is always less than 10 \%
for the values of the anomalous couplings chosen in Figure 1.
For  somewhat larger anomalous couplings, say
$(\delta \kappa, \lambda)=(2.5,2.5)$, we find that $R_{hard}$ can be
as large as  7 and 24 \% for a 500 GeV and 1 TeV Higgs
respectively using the same photon energy cutoff.  As is evident in Figure 1,
destructive effects occur mainly for a positive
$\delta \kappa$ and a negative
$\lambda$. In other cases, the anomalous contributions tend to have
constructive interference with the SM result as the Higgs mass grows heavier.
Increasing (decreasing)  the photon energy cutoff tends
to decrease (increase) the branching ratio.
We also see that $R_{hard}$ increases monotonically with the Higgs mass
when all the other parameters are held fixed.
For $(\delta \kappa, \lambda)=(0.5,0.5)$ and (1,1),
$R_{hard}$ approaches to 100  \% as the Higgs mass becomes
10 and 5 TeV respectively.
On the other hand, $R_{hard}$ climbs up to about 22 \%
for a 10 TeV Higgs in the SM  with $(\delta \kappa, \lambda)=(0,0)$.
At any rate, perturbative calculation is no longer trustworthy
for such a heavy Higgs.

To conclude, we have studied in detail the
radiative decay mode of a heavy Higgs
$H \rightarrow W^+W^-\gamma$ for a {\it hard} photon in a wide class of model
(including the SM) parametrized  by the anomalous couplings $\kappa$ and
$\lambda$. The SM prediction for the branching ratio $R_{hard}$ is only a
few percent and the anomalous contributions tend to increase its value
somewhat but never exceeds 10 percent unless the magnitudes of the anomalous
couplings turn out to be much larger than unity or the Higgs boson becomes
ultra-heavy.

\vfill

\newpage

\acknowledgments

I would like to thank Professor W.--Y. Keung for bringing the Standard Model
calculation (Ref.\cite{dicusetal}) to my attention.
This work was supported by the
Department of Energy under contracts DE--AC02--76-ER022789.

\vfill

\newpage

\unletteredappendix{ }

In this Appendix, we collect the analytic results for all the integrals
defined in the branching ratio $R_{hard}$. One can split the total
contribution  into the SM piece and an extra piece
arise from the anomalous couplings,
\begin{equation}
R_{hard}  \; = \; {2\alpha \over \pi }(1-r^2)^{-{1 \over 2}}
       \left[ C_{SM} + \delta C \right] \;\;\; .
\end{equation}
The SM contribution is given by\cite{comment}
\begin{equation}
C_{SM} \; = \; (1-{r^2 \over 2})A_1 - A_2 - B_1 + 8(3r^4-4r^2+4)^{-1}B_5
\; \; ,
\end{equation}
and the anomalous piece is
\begin{equation}
\begin{array}{cl}
\delta C & =  \; (3r^4-4r^2+4)^{-1} \\
&  \times \left\{  2 \lambda \left[ 2(A_3 - B_2 + B_4) + r^2A_3 \right]
     - 2 \delta \kappa \left[ 2(A_3 - B_2 + B_4 - 2B_5) - r^2A_3 \right]
\right.
 \\
&   + {\lambda^2 \over 6r^2} \left[ 4(B_2 - B_4 + 2B_5 - 7B_6)
     + r^2(B_2 -B_3 -2B_4 -B_5 - 6A_3 + 18A_4) + 6r^4A_3 \right] \\
&   + {(\delta \kappa)^2 \over 6r^2} \left[ 4(B_2 - B_4 + 2B_5 - B_6)
     + r^2(B_2 -B_3 -2B_4 -B_5 - 6A_3 - 6A_4) + 6r^4A_3 \right] \\
&   \left. - {\lambda \delta \kappa \over 3r^2}
      \left[ 4(B_2 - B_4 - 4B_5 + 2B_6)
     + r^2(B_2 -B_3 -2B_4 -B_5 + 6A_3 - 6A_4) - 6r^4A_3 \right]
\right\}  . \\
\end{array}
\end{equation}
$A_i(i=1 \;  {\rm to} \; 4)$ and $B_i(i=1 \; {\rm to} \; 6)$ are the integrals
defined by
\begin{equation}
A_{1,2,3,4} \; = \; \int_y^{1-r^2} dx \, \left( {1 \over x},1,x,x^2 \right)
            {\rm ln} \left[ {1-x+R(x,r) \over 1-x-R(x,r)} \right] \;\; ,
\end{equation}
\begin{equation}
B_{1,2,3,4,5,6} \; = \;
  \int_y^{1-r^2} dx \, \left( {1 \over x},{1 \over 1-x},{1 \over (1-x)^2},
1,x,x^2 \right) R(x,r) \;\; ,
\end{equation}
where $R(x,r)$ was defined in Eq.(6). Evaluating these integrals
are laborious. The final results are
\begin{equation}
\begin{array}{cl}
A_1 \; = & - \ln^2 2 + \ln \, y \, \ln \, r^2 \,
  + \ln^2 \left( {1-y-R(y,r) \over r^2} \right) - 2 \, \ln \, 2 \,
        \ln \left( {1-y+R(y,r) \over r^2} \right) \\
& + 2 \, \ln \, (1+ \sqrt{1-r^2}) \, \ln \, \left(
  { (1-r^2)(1-y)+\sqrt{1-r^2}R(y,r) \over r^2y } \right) \\
& + 2 \, \ln \, (1- \sqrt{1-r^2}) \, \ln \, \left(
  { (1-r^2)(1-y)-\sqrt{1-r^2}R(y,r) \over r^2y } \right) \\
& +2 \, \li \left( { r^2(1-y)-(1-\sqrt{1-r^2})(1-y+R(y,r))
            \over r^2(1-y) } \right)
  +2 \, \li \left( { r^2(1-y)-(1+\sqrt{1-r^2})(1-y+R(y,r))
            \over r^2(1-y) } \right) \\
& -2 \, \li \left( {-1+r^2+\sqrt{1-r^2} \over r^2} \right)
  -2 \, \li \left( {-1+r^2-\sqrt{1-r^2} \over r^2} \right) \\
& + \li (y) + \li(r^2)
            - 2 \, \li \left(
            {1-y-R(y,r) \over 2(1-y)} \right) \;\; , \\
\end{array}
\end{equation}
\begin{equation}
A_2 \; = \; - R(y,r) + (1-y-{r^2 \over 2}) \, L(y,r) \;\;,
\end{equation}
\begin{equation}
A_3 \; = \; {1 \over 8} \left[ 3r^2-8 +2(1-y) \right] R(y,r)
 + {1\over 16} \left[ r^2(3r^2-8) + 8(1-y^2) \right] \, L(y,r)
\;\;,
\end{equation}
\begin{equation}
\begin{array}{cl}
A_4 \; = & {1 \over 72} \left[ r^2(44+10y-15r^2)-4(11+5y+2y^2)
          \right] R(y,r) \\
 & - {1\over 48} \left[ r^2(5r^4-18r^2+24) - 16(1-y^3) \right]
  \, L(y,r) \;\;, \\
\end{array}
\end{equation}
\begin{equation}
B_1 \; = \; -R(y,r) + {1 \over 2}(r^2-2) \, L(y,r)
 +(1-r^2)^{1 \over 2} \, L'(y,r)
\;\;,
\end{equation}
\begin{equation}
B_2 \; = \; R(y,r) - {r^2 \over 2} \, L(y,r) \;\;,
\end{equation}
\begin{equation}
B_3 \; = \; -{2 \over 1-y}R(y,r) + \, L(y,r) \;\;,
\end{equation}
\begin{equation}
B_4 \; = \; - {1 \over 4}\left[ r^2 -2(1-y) \right] R(y,r)
        - {r^4 \over 8} \, L(y,r) \;\;,
\end{equation}
\begin{equation}
B_5 \; = \; -{1 \over 24}\left[ r^2(2(2+y)-3r^2) -4(1+y(1-2y)) \right] R(y,r)
        + {r^4(r^2-2) \over 16} \, L(y,r) \;\;,
\end{equation}
\begin{equation}
\begin{array}{cl}
B_6 \; = & - {1 \over 192} \left[ 15r^6 - 2r^4(19+5y) + 8r^2(3+2y+y^2)
                 - 16(1+y+y^2-3y^3) \right] R(y,r) \\
   &     - {r^4 \over 128}(5r^4 -16r^2 +16) \, L(y,r) \;\; .
 \\
\end{array}
\end{equation}
In the above equations, we have defined
\begin{equation}
\begin{array}{cl}
L(y,r) \; & = \; \ln \, \left( {2(1-y)-r^2+2R(y,r) \over r^2} \right)
\; \; , \\
L'(y,r) \; &  = \; \ln \, \left( {2(1-r^2)+(r^2-2)y+2(1-r^2)^{1 \over 2}
 R(y,r) \over r^2y} \right) \;\; . \\
\end{array}
\end{equation}

\figure{$R_{hard}$ as function of Higgs mass with $E^{min}_{\gamma}$ = 10 GeV.
(a) $(\delta \kappa , \lambda) = (\pm 0.5, \pm 0.5)$ and
(b) $(\delta \kappa , \lambda) = (\pm 1, \pm 1)$. The Standard Model
prediction ($\delta \kappa =0, \lambda = 0$) is also presented
for comparison. We take $w$ = 80 GeV and $\alpha$ = 1/128.}


\begin{references}
\bibitem{comment}
The result of $C_{SM}$ in Eq.(A2) agrees with Ref.[1]. However, the formula
of $C_{SM}$ given in Ref.[1] is only semi-analytic since the integral
$A_1$ was evaluated using the mean valued theorem in calculus. We have
checked  their semi-analytic result agrees well with our analytic result
using Eq.(A6) of $A_1$.
\end{references}
\end{document}